\documentclass[sigconf]{acmart}

\acmConference[]{Conference}{}{2026}

\usepackage{pifont}
\usepackage[dvipsnames]{xcolor}
\usepackage{array}
\usepackage{xcolor,soul}
\usepackage{tikz}
\usepackage{pgf-pie}
\usepackage{xspace}
\usepackage{cleveref}
\usepackage{subcaption}
\usepackage[T1]{fontenc}
\usepackage[utf8]{inputenc}
\usepackage{listings}
\usepackage{xcolor}
\usepackage{soul}
\usepackage{tcolorbox}
\usepackage{multirow} 
\usepackage{enumitem}

\lstdefinelanguage{pseudocode}{
  morekeywords={procedure,for,each,while,receive,match,case,if,else,end},
  sensitive=true,
}

\usetikzlibrary{arrows.meta}

\usepackage{mdframed}

\definecolor{ao}{rgb}{0.12, 0.3, 0.17}
\definecolor{darklavender}{rgb}{0.45, 0.31, 0.59}

\newcommand{\eclipsecolor}[1]{\textcolor{violet}{#1}}
\newcommand{\approach}{GORE\xspace}
\newcommand{\boldparagraph}[1]{\vskip 0.05in\noindent\textbf{#1.}}
\newcommand{\sleec}[1]{{\small\fontfamily{qcr}\selectfont#1}}

\newcommand{\sleeckeyword}[1]{\eclipsecolor{\textbf{#1}}}

\AtBeginDocument{%
  }

\begin{document}

\title{Operationalizing Human Values in the Requirements Engineering Process of Ethics-Aware Autonomous Systems}

\author{Everaldo Silva Junior}
\affiliation{
  \institution{University of Brasilia}
  \country{Brazil}
}
\email{junior.everaldo@aluno.unb.br}

\author{Lina Marsso}
\affiliation{
    \institution{Polytechnique Montreal}
    \country{Canada}
}
\email{lina.marsso@polymtl.ca}

\author{Ricardo Caldas}
\affiliation{
    \institution{Gran Sasso Science Institute}
    \country{Italy}
}
\email{ricardo.caldas@gssi.it}

\author{Marsha Chechik}
\affiliation{
    \institution{University of Toronto}
    \country{Canada}
}
\email{chechik@cs.toronto.edu}

\author{Genaína Nunes Rodrigues}
\affiliation{
  \institution{University of Brasilia}
  \country{Brazil}
}
\email{genaina@unb.br}

\begin{abstract}
Operationalizing human values alongside functional and adaptation requirements remains challenging due to their ambiguous, pluralistic, and context-dependent nature. Explicit representations are needed to support the elicitation, analysis, and negotiation of value conflicts beyond traditional software engineering abstractions. In this work, we propose a requirements engineering approach for ethics-aware autonomous systems that captures human values as normative goals and aligns them with functional and adaptation goals. These goals are systematically operationalized into Social, Legal, Ethical, Empathetic, and Cultural (SLEEC) requirements, enabling automated well-formedness checking, conflict detection, and early design-time negotiation. We demonstrate the feasibility of the approach through a medical Body Sensor Network case study.
\end{abstract}




\maketitle
\pagestyle{plain}

\section{Introduction} 
Digital humanism provides an ethical foundation for the digital age by framing digital transformation around humanist principles such as authorship, responsibility, reason, and freedom~\cite{DigitalHumanismBook:2024}. As software increasingly mediates societal functions, requirements engineering (RE) becomes a central locus of responsibility, shaping not only system behavior but also the values systems embody and how those values are negotiated over time. 
Prior work has proposed methods for incorporating human values into RE~\cite{Bennaceur:SEAMS2023, CURUMSING:JSS2019, Sarah:RE2018, Shahin:2022, Bennaceur2024:ResponsibleSoftwareEngineering}. However, the systematic operationalization of values alongside functional and adaptation requirements across the RE lifecycle remains challenging, particularly due to the ambiguous, pluralistic, and context-dependent nature of human values~\cite{ALIDOOSTI2025112430}. Explicit representations are still needed to elicit, model, analyze, and negotiate value conflicts in ways that go beyond traditional software engineering abstractions~\cite{Shahin:2022}.

In autonomous systems, these challenges are amplified by deployment in safety- and society-critical domains such as healthcare, robotics, and autonomous driving. Such systems must continuously adapt at runtime, requiring precise and analyzable requirements, often captured through goal-based models, to derive control strategies~\cite{Filieri:2015, 8500345, 10496502, 10164693, BSN:SEAMS2018}. Despite advances in adaptive and goal-oriented systems, ethical values remain weakly integrated into the RE pipelines that drive adaptation.
This gap raises a central challenge: how can pluralistic and evolving human values be translated into requirements that are precise enough to guide adaptation decisions of autonomous systems, while remaining robust to contextual change? Addressing this challenge also opens opportunities across the RE lifecycle, from explicit elicitation of value trade-offs, to goal-based modeling with traceability, systematic conflict analysis, and accountable negotiation of ethical decisions.

Recent work on the LEGOS-SLEEC framework~\cite{LEGOS:ICSE25} has advanced the operationalization of social, legal, ethical, empathetic, and cultural (SLEEC) values through automated well-formedness and consistency checking of rule-based requirements. While effective for identifying inconsistencies and supporting rule-level debugging, SLEEC-based approaches primarily operate at the rule level, leaving stakeholder intent, links to functional requirements, and adaptation mechanisms implicit.

In this paper, we address these limitations by integrating goal-oriented requirements engineering with the rule-based SLEEC approach to support ethics-aware adaptation. Our approach makes stakeholder intent explicit through goals while preserving SLEEC’s analytical strengths, such as consistency and well-formedness checking. By establishing an explicit contract with stakeholders and future users~\cite{DigitalHumanismBook:2024}, it supports early feedback and traceable reasoning during the engineering of discrete-event self-adaptive controllers. 

Concretely, we contribute:
(i) a goal-oriented value model in which ethical values are represented as normative goals aligned with functional and adaptation goals;
(ii) structured analysis and well-formedness checking of relationships among values, system behavior, and adaptation goals, including conflict detection;
(iii) an integrated RE framework unifying functional, non-functional, and ethical requirements to reason explicitly about trade-offs; and
(iv) support for preserving ethics awareness during design and adaptation through traceability from stakeholder values to operational requirements and adaptation decisions. 

We demonstrate the feasibility of
the approach through a medical Body Sensor Network case study~\cite{BSN:SEAMS2018}. Our results show that our approach is effective in operationalizing human values in ethics-aware autonomous systems by systematically refining high-level requirements into analyzable goals and tasks, which can be automatically translated into SLEEC rules using our LEGOS-SLEEC extension.
\section{Background}\label{sec:background}
\boldparagraph{SLEEC DSL}
The main concepts of the \sleec{SLEEC} DSL~\cite{Getir-Yaman-et-al-23} consist of definitions and rules (see Tbl.~\ref{tab:sleecRules}). 
Definitions declare events and measures representing system capabilities and its activities during the interaction with the environment, including humans. 
\sleeckeyword{Events} represent instantaneous actions whereas \sleeckeyword{measures} represent capabilities to provide (immediately) information captured by values of data types, such as \sleeckeyword{Boolean}, \sleeckeyword{numeric}, and \sleeckeyword{scale}. In the SLEEC DSL, events are capitalized while measures are not. 
Rules have the basic form ``\sleec{\sleeckeyword{when} trigger \sleeckeyword{then} response}''. Such a rule defines the required response when the event in the trigger happens and its conditions on measures, if any, are satisfied. For example, rule \sleec{r2} applies when the event \sleec{AdaptationExecuted} occurs, in which case the response \sleec{ExplainAdaptation} is required within the next $2$ minutes.  
A SLEEC rule can be accompanied by one or more \emph{defeaters}, introduced using the ``\sleec{\sleeckeyword{unless}}'' construct. The language incorporates time constructs allowing responses with deadlines and timeouts using the ``\sleec{\sleeckeyword{within}}'' construct, as seen in rule \sleec{r2}. 

\begin{table*}[htb!]
    \centering
    \caption{{\small Normative requirements for the Body Sensor Network in SLEEC DSL.}} 
    \vspace{-0.1in}
    \label{tab:sleecRules}

    \vspace{-0.1in}
    \scalebox{0.75}{
        \begin{tabular}{r  l}
        \\
        \toprule
        \multicolumn{2}{c}{\sleec{\textbf{Definitions}}}\\
        \midrule
             \sleec{\sleeckeyword{event}} & \sleec{UserAsksStopTracking}\\
             \sleec{\sleeckeyword{event}} & \sleec{StopTracking}\\
             \sleec{\sleeckeyword{event}} & \sleec{CallCaregiver}\\
             \sleec{\sleeckeyword{event}} & \sleec{AdaptationExecuted}\\
             \sleec{\sleeckeyword{event}} & \sleec{ExplainAdaptation}\\
             \sleec{\sleeckeyword{event}} & \sleec{UserRequestsPrivacy}\\
             \sleec{\sleeckeyword{measure}} & \sleec{trackVitals: \sleeckeyword{boolean}} \\
             \sleec{\sleeckeyword{measure}} & \sleec{riskLevel: \sleeckeyword{scale}(low, medium, high)} \\
        \bottomrule
    \end{tabular}
    }
    \scalebox{0.76}{
        \begin{tabular}{rl}
        \\
        \toprule
        \multicolumn{2}{c}{\sleec{\textbf{Rules}}}\\
        \midrule
             \sleec{r1:=} & \sleec{\sleeckeyword{when} UserAsksStopTracking \sleeckeyword{and} {trackVital} \sleeckeyword{then}} \\
             & \sleec{StopTracking \sleeckeyword{within} 5 \sleeckeyword{minutes}} \\
             & \sleec{\sleeckeyword{unless} (riskLevel > medium) \sleeckeyword{then}} \\
             & \sleec{CallCaregiver  \sleeckeyword{within} 5 \sleeckeyword{minutes}}\\
             \sleec{r2:=} &  \sleec{\sleeckeyword{when} AdaptationExecuted \sleeckeyword{then}} \\ & \sleec{ExplainAdaptation \sleeckeyword{within} 2 \sleeckeyword{minutes}}\\
             \sleec{r3:=} & \sleec{\sleeckeyword{when} UserRequestsPrivacy \sleeckeyword{and} (\sleeckeyword{not} userConsent)} \\
             & \sleec{\sleeckeyword{then} \sleeckeyword{not} CallCaregiver \sleeckeyword{within} 5 \sleeckeyword{minutes}} \\
        \bottomrule
        \end{tabular}
    }
\end{table*}

\boldparagraph{Well-formedness properties} 
\textit{(1) Vacuous conflicts.}
A rule is \emph{vacuously conflicting} if its trigger is not part of the accepted behaviours defined by the rule set, meaning that triggering the rule in any situation inevitably leads to a violation of another rule. 
\textit{(2) Situational conflicts.}
A rule $r$ is \emph{situationally conflicting} if there exists a feasible situation in which $r$ is triggered and its required response conflicts with the response of another applicable rule.  
\textit{Example}. consider the \sleec{SLEEC} rules \sleec{r1} and \sleec{r3} shown in Tbl.~\ref{tab:sleecRules}. \sleec{r1} conflicts with \sleec{r3} in a situation where the user requests to stop vital-sign monitoring while the assessed risk level is high, and the user simultaneously requests privacy without consenting to escalation. This results in \sleec{r3} blocking the caregiver notification required by \sleec{r1}, yielding a situational conflict that reflects a trade-off between non-maleficence and respect for user autonomy and privacy.
\section{\approach: Approach for Ethics-Aware Adaptation}~\label{sec:methodology}
\begin{figure*}[htb!]
  \centering
\includegraphics[width=.9\linewidth]{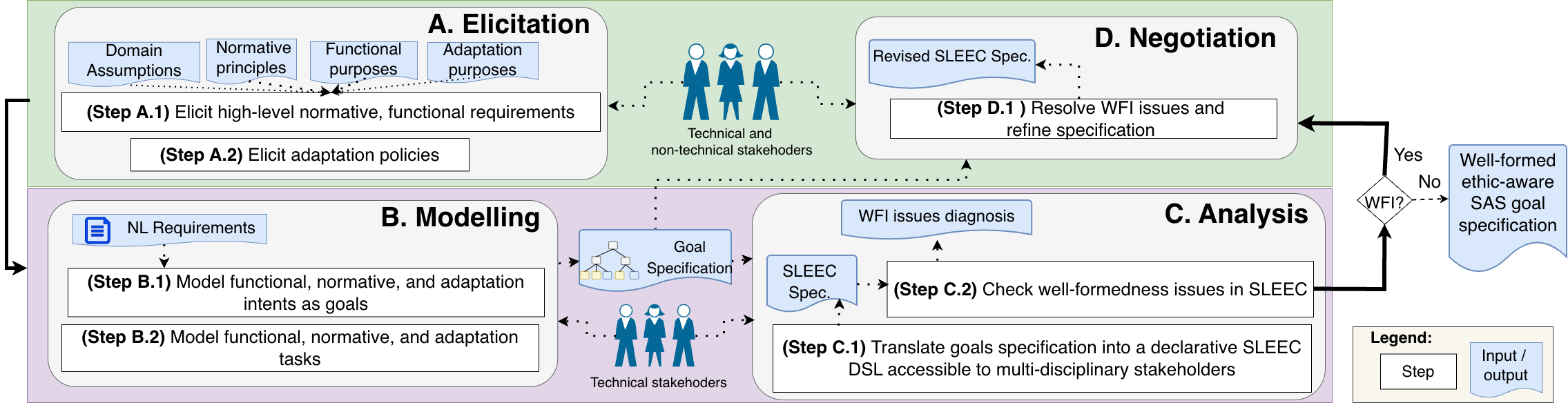}
  \vspace{-0.1in}
  \caption{Operationalization of Human Values in the Requirements Engineering Process of Ethics-Aware Autonomous Systems.}
  \Description{}
  \label{fig:mainApproach}
  \vspace{-0.1in}
\end{figure*}

Incorporating human values into software engineering requires rethinking requirements around evolving stakeholders, their needs, and values, while acknowledging inevitable change and failure. This is challenging because values are subjective and context-dependent: stakeholders may interpret the same value differently, and its implications often emerge only after deployment through real-world use and unforeseen trade-offs. We treat normative principles as first-class entities~\citet{Bennaceur2024:ResponsibleSoftwareEngineering} and operationalize them alongside functional and adaptation requirements. Our approach, \approach{} (Fig.~\ref{fig:mainApproach}), is an iterative process structured around four activities: (A) elicitation, (B) modeling, (C) analysis, and (D) negotiation.

\subsection{Elicitation}
Ensuring ethical decision-making in software systems has engaged social scientists, ethicists, policymakers, technologists, and civil society. Participatory design supports this effort by eliciting values from diverse stakeholders through user involvement, personas and prototypes, iterative feedback, and attention to empathy and cultural diversity~\cite{ORBIT_RRI_AREA4P, Mackay2020, Peng2022DesignThinking, ANDERSONCOTO2024100621}. These practices make stakeholder values explicit and actionable throughout the design process~\cite{Bennaceur2024:ResponsibleSoftwareEngineering}. In the elicitation activity, normative principles, functional and adaptation purposes, and domain assumptions guide stakeholders in identifying high-level requirements, which are then refined into explicit requirements.


\textbf{\underline{Step A.1}: Elicit high-level requirements.} 
This step elicits functional and normative requirements with stakeholders, expressed in natural language. Functional requirements are derived using established requirements engineering techniques (e.g., interviews, workshops, scenarios, and use cases) to capture system objectives, expected services, and domain assumptions~\cite{Lamsweerde:REBook}. Normative requirements are elicited using a pluralistic approach~\cite{Townsend2022Pluralistic}, which engages diverse stakeholders to examine system capabilities and domain assumptions and identify relevant ethical, legal, social, empathetic, and cultural principles. These principles are translated into actionable proxies and expressed as structured normative requirements associated with system capabilities and domain assumptions. 
\textit{\underline{Example}.} To illustrate our approach, we use the self-adaptive body sensor network (SA-BSN)~\cite{BSN:SEAMS2021}, which detects emergencies through continuous health monitoring and adapts its behavior to changes in patient condition and consent. For a Body Sensor Network (BSN) for health monitoring, functional requirements include periodic vital-sign collection, data fusion by a central hub, and health-risk detection. Normative elicitation focuses on patient consent for partial or full monitoring. Given the sensitivity of health data, privacy and autonomy emerge as primary ethical principles, alongside legal compliance with healthcare and data-protection regulations. These principles define obligations to respect informed patient choices, restrict monitoring to authorized measurements, and promptly enforce consent changes, forming the basis for consent-related requirements.

\textbf{\underline{Step A.2}: Elicit adaptation policies.}
This step elicits adaptation policies from adaptation purposes and ethical concerns, focusing on what the system should adapt, under which conditions, and which values must be preserved—independently of implementation mechanisms. This separation enables early reasoning about correctness, compliance, and ethical trade-offs before committing to architectural or control solutions. Following requirements-driven approaches to self-adaptive systems~\cite{WeynsBook:Ch07}, adaptation concerns are captured as first-class requirements and operationalized as adaptation goals. Because self-adaptive systems operate under uncertainty, elicited policies must explicitly account for variability, conditions, and adaptation logic. 
\textit{\underline{Example}.} In the BSN example, adaptation is driven by changes in patient consent (e.g., switching from full to partial monitoring). At elicitation, the adaptation policy is captured as an intentional relationship between normative, adaptation, and functional purposes, specifying when and how monitoring behavior must change and which vital signs may be collected. This policy captures stakeholder expectations prior to modeling or implementation decisions.

\subsection{Modelling}
%
We adopt the dual interpretation of normative requirements as both functional and non-functional~\cite{CURUMSING:JSS2019}, since they express hard goals the system must satisfy and soft goals capturing qualitative concerns such as privacy, autonomy, or discomfort~\cite{Townsend2022Pluralistic}. To make this dual nature explicit~\citet{Detweiler:AOSE11}, we introduce a dedicated modeling notation for normative principles. This notation links system goals to underlying values and supports the identification of value trade-offs and conflicts during analysis. 
Building on this foundation, the modeling phase structures elicited requirements into a goal-oriented model. In Step B.1, normative, functional, and adaptation goals capture what the system should achieve and which values it must uphold. In Step B.2, these goals are operationalized into tasks describing how they are realized at runtime. The resulting model explicitly connects values, goals, and operationalizations, forming the basis for analysis and adaptation reasoning.

\textbf{\underline{Step B.1}: Model functional, normative, and adaptation intents as goals.}
In our approach, goals capture the intents of stakeholders regarding a system, reflecting the desired impacts on their lives. Since stakeholders often have diverse and sometimes conflicting goals, it is crucial to identify and model these goals clearly to facilitate future trade-off decisions \cite{Letier2025_RE_Guide}. 
In our modeling approach, we distinguish three types of goals. 
\emph{Normative goals} capture ethical, social, and empathetic properties that guide system behavior and ensure alignment with human values. 
\emph{Functional goals} describe the non-normative behaviors the system must exhibit to fulfill its intended functionality. 
\emph{Adaptation goals} specify how the system should adjust its configuration or behavior in response to changes in order to support self-adaptation. 
\Cref{tab:functionalgoal}\footnote{While functional and adaptation goals share the same structure, adaptation goals further constrain system reactions to context changes. We defer details to an extended version.} presents the common attributes for functional and adaptation goals, while \Cref{tab:normativegoal} presents those of normative goals. 
\textit{\underline{Example}.} 
\begin{figure}[htb!]
    \centering
    \includegraphics[width=0.8\linewidth]{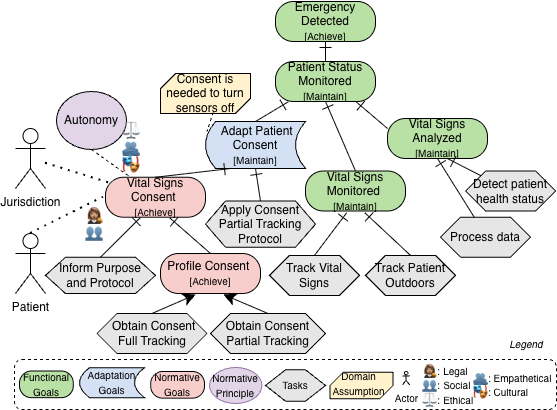}
    \caption{Goal Model for the Body Sensor Network System.}
    \label{fig:BSN}
\end{figure} 
In Fig.~\ref{fig:BSN}, {\tt Vital Signs Consent} is a normative goal representing the granting and withdrawal of permissions (e.g., consent and assent), ensuring user control~\cite{Townsend2022Pluralistic}. It aligns with the human value of autonomy and applies during system configuration, calibration, and whenever the user updates their monitoring profile. We next map this scenario to the attributes of our normative goal, as presented in \Cref{tab:normativegoal:partialconsent}.
\begin{table}[t]
\caption{{\small Common attributes for functional and adaptation goal}}
\vspace{-0.1in}
\label{tab:functionalgoal}
\centering
\small
\scalebox{0.8}{
    \begin{tabular}{p{2cm}p{6cm}}
    \toprule
    \textbf{\scshape Attribute} & \textbf{Description} \\
    \midrule
    \scshape Goal & \textit{<name>} \\
    \scshape Type & maintain \;|\; achieve \\
    \scshape Condition & state to maintain or achieve \\
    \scshape Event & good or target event \\
    \scshape ContextEvent & current or triggering event \\
    \scshape Def & natural language description \\
    \scshape FormalDef & formal specification \\
    \bottomrule
    \end{tabular}
}
\end{table}

\begin{table}[htb!]
\caption{{\small Normative goal specification attributes.}}
\vspace{-0.1in}
\label{tab:normativegoal}
\centering
\small
\scalebox{0.8}{
    \begin{tabular}{p{2cm}p{6cm}}
    \toprule
    \textbf{\scshape Attribute} & \textbf{Description} \\
    \midrule
    \scshape NORMATIVEGoal & \textit{<name>} \\
    \scshape Type & maintain \;|\; achieve \\
    \scshape Source & laws \;|\; regulations \;|\; standards \;|\; protocols \\
    \scshape Class & Social \;|\; Legal \;|\; Ethical \;|\; Empathetic \;|\; Cultural \\
    \scshape NormPrinciple & Autonomy \;|\; Privacy \;|\; Transparency \\
    \scshape Proxy & Consent, Disclosure, etc. \\
    \scshape AddedValue & value contribution \\
    \scshape Condition & state to maintain or achieve \\
    \scshape Event & good or target event \\
    \scshape ContextEvent & current or triggering event \\
    \scshape Def & natural language description \\
    \scshape FormalDef & formal specification \\
    \bottomrule
    \end{tabular}
}
\end{table}

\begin{table}[htb!]
\caption{{\small Vital Signs Consent specification attributes}}
\vspace{-0.1in}
\label{tab:normativegoal:partialconsent}
\centering
\small
\scalebox{0.8}{
    \begin{tabular}{p{2cm}p{7.8cm}}
    \toprule
    \textbf{\scshape Attribute} & \textbf{Description} \\
    \midrule
    \scshape NORMATIVEGoal & Vital Signs Consent \\
    \scshape Type & achieve \\
    \scshape Source & Healthcare and data-protection regulations; medical ethics \\
    \scshape Class & Ethical; Legal; Social \\
    \scshape NormPrinciple & Autonomy \\
    \scshape Proxy & Assent/Consent \\
    \scshape AddedValue & Preserved patient autonomy and reduced privacy intrusion \\
    \scshape Condition & purposeProtocolInformed = true \& patientConsentsFullTracking = true \\
    \scshape Event & AchievedObtainConsentFullTracking \\
    \scshape ContextEvent & MeetingUser \\
    \scshape Def & Restrict monitoring to patient-authorized vital signs only \\
    \scshape FormalDef &
    $\forall s \in Sensors:$ $(ConsentStatus = Full \wedge s \notin AuthorizedSensors)$
     $\Rightarrow \neg Monitor(s)$ \\
    \bottomrule
    \end{tabular}
}
\end{table}

\textbf{\underline{Step B.2}: Model functional, normative, and adaptation tasks.} 
A task represents a desired behavioral property of the machine that must be performed to satisfy stakeholder goals\cite{Letier2025_RE_Guide, Lamsweerde:REBook}. Task modeling involves understanding these goals and defining how they can be operationalized while considering the system's capabilities. Following a top-down strategy, goals are systematically decomposed into tasks, with relevant domain assumptions captured as preconditions or triggering events. Tasks follow the notation in Tab.~\ref{tab:normativetask}. 
\textit{\underline{Example}.} As illustrated in~\Cref{fig:BSN}, the tasks of the BSN are shaded in gray. Examples of tasks include: \emph{Inform Purpose and Protocol}, \emph{Obtain Consent Full Tracking}, or \emph{Apply Consent Partial Tracking Protocol}. 
Each task defines operations executed to fulfill the BSN's functional, adaptation, and normative goals.

\begin{table}[htb!]
\caption{{\small Normative task specification attributes.}}
\vspace{-0.1in}
\label{tab:normativetask}
\centering
\small
\scalebox{0.8}{
    \begin{tabular}{p{2.5cm}p{6cm}}
    \toprule
    \textbf{\scshape attribute} & \textbf{Description} \\
    \midrule
    \scshape Task & \textit{<NormativeTask>} \\
    \scshape Def & Natural language definition of the normative task. \\
    \scshape PreCond & Required system or environmental state. \\
    \scshape TriggeringEvent & Event that initiates task execution. \\
    \scshape TemporalConstraint & Time constraint (e.g., deadline, duration). \\
    \scshape PostCond & State resulting from satisfying the normative task. \\
    \scshape ObstacleEvent & Event that inhibits or restricts task execution. \\
    \bottomrule
    \end{tabular}
}
\vspace{-0.08in}
\end{table}

\subsection{Analysis}
Values are inherently subjective, making conflicts and violations difficult to anticipate and monitor. Prior work in value-based requirements engineering proposes methods to surface value conflicts through iterative elicitation and refinement of stakeholders' values and motivations~\cite{Sarah:RE2018, Bennaceur:SEAMS2023, Shahin:2022}. However, systematic mechanisms for analyzing and resolving such conflicts remain limited~\cite{Shahin:2022}. Our approach addresses this gap by leveraging the SLEEC DSL and its satisfiability checking process to detect well-formedness issues in requirements. Steps C.1 and C.2 describe this process.

\textbf{\underline{Step C.1}: Translate goals specification into a declarative SLEEC DSL accessible to multidisciplinary stakeholders}.  
Requirements elicited and modeled as functional, normative, and adaptation goals provide a structured and analyzable representation of system objectives, but are not easily accessible to non-technical stakeholders. Since philosophers, lawyers, ethicists, medical practitioners, and other domain experts primarily work with natural-language artifacts, this step translates goal models into a declarative specification expressed in the SLEEC DSL (Sec.~\ref{sec:background}). SLEEC remains close to natural language while providing precise semantics, enabling stakeholder review, validation, and automated analysis.

\textit{Automated translation from goals to SLEEC specifications.} 
Translation is performed automatically using predefined transformation rules that bridge goal models and the event-driven semantics of SLEEC. This bridge relies on \emph{fluents}~\cite{Letier:Fluents2005}, which discretize system behavior by capturing task lifecycles through initiating and terminating events. Each task is represented as a fluent initially set to false, formally: $\mathit{Fluent} \text{ Task} = \langle \mathit{\{StartTask\}}, \mathit{\{AchievedTask\}} \rangle \mathit{initially} \textbf{ false}$ 
Initiating events mark task execution, while terminating events denote completion. This representation enables traceable task execution and the systematic derivation of SLEEC rules from goal models. Table~\ref{tab:mappingtemplates} summarizes the mapping between goal constructs and SLEEC DSL patterns.

\begin{table}[htb!]
  \caption{{\small Goal model to SLEEC DSL mapping templates.}}
  \vspace{-0.1in}
  \label{tab:mappingtemplates}
  \centering
  \setlength{\tabcolsep}{4pt} 
  \scalebox{0.77}{
    \begin{tabular}{p{1.9cm}p{8.2cm}}
      \toprule
      \textbf{Goal element} & \textbf{SLEEC DSL template} \\
      \midrule
      Maintain Goal &
      \sleec{P1 \sleeckeyword{exists} Event \sleeckeyword{and} Condition \sleeckeyword{while} ContextEvent} \\\hline

      Achieve Goal &
      \sleec{P2 \sleeckeyword{when} ContextEvent \sleeckeyword{and} Condition \sleeckeyword{then} Event} \\\hline

      \multirow{4}{*}{Task} & \sleec{T1 \sleeckeyword{when} TriggeringEvent \sleeckeyword{and} PreCond \sleeckeyword{then} StartTask} \\
      \cline{2-2}
                             & \sleec{T2 \sleeckeyword{when} StartTask \sleeckeyword{then} PursuingTask \sleeckeyword{within} TemporalConstraint} \\
      \cline{2-2}
                             & \sleec{T3 \sleeckeyword{when} PursuingTask \sleeckeyword{and} PostCond \sleeckeyword{then} AchievedTask \sleeckeyword{unless not} PostCond \sleeckeyword{then} ReportFailureTask} \\
      \cline{2-2}
                             & \sleec{Obstacle \sleeckeyword{when} ObstacleEvent \sleeckeyword{then not} PursuingTask} \\
      \bottomrule
    \end{tabular}
  }
\end{table}

\begin{figure}[htb!]
    \centering
    \includegraphics[width=.9\linewidth]{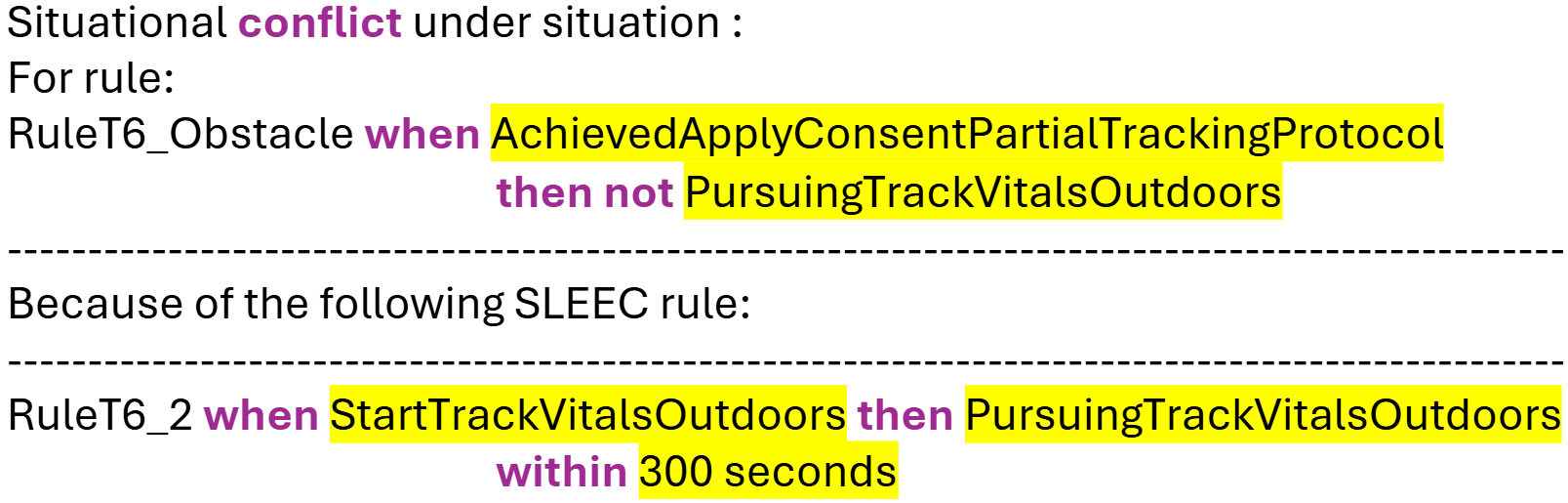}
    \vspace{-0.1in}
    \caption{{\small Situational conflict diagnosis.}}
    \label{fig:situational}
\end{figure}

\textbf{\underline{Step C.2}: Check well-formedness issues in SLEEC specifications.}
At this stage, functional, normative, and adaptation requirements have been translated into SLEEC rules. Step~C.2 analyzes their well-formedness to assess achievability using the LEGOS-SLEEC tool~\cite{NormativeRequirements:ICSE2024}. The tool detects rule conflicts, both globally and in specific adaptation contexts, and provides diagnostic feedback identifying the conflicting rules and enabling traceability to the underlying values and trade-offs elicited earlier. 
\textit{\underline{Example}.} 
Using LEGOS-SLEEC, we obtained the diagnosis shown in Fig.~\ref{fig:situational}, which highlights the root cause of a situational conflict between \textit{Apply Consent Partial-Tracking Protocol} and \textit{Track Patient Outdoors}, corresponding to RuleT6\_Obstacle and RuleT6\_2. These tasks operationalize the adaptation goal \textit{Adjust Deployed Consent} and the functional goal \textit{Vital Signs Monitored}. The conflict arises when a new consent profile is deployed while outdoor monitoring is still executing: the updated consent invalidates the ongoing activity, yet interrupting it would prevent satisfaction of the monitoring goal. As a result, the controller reaches a blocking situation.

\subsection{Negotiation}\label{sec:negotiation}

In ethics-aware autonomous systems, negotiation makes trade-offs among functional, normative, and adaptation goals explicit, justified, and revisable through stakeholder agreement. This negotiation is grounded in ethical and regulatory constraints that govern system behavior under changing conditions. A key distinction is between \emph{hard ethics}, which impose non-negotiable constraints such as safety and certification, and \emph{soft ethics}, encompassing values such as autonomy, privacy, and fairness, which admit contextual interpretation and controlled trade-offs~\cite{Floridi2018SoftEthics}. While adaptation mechanisms may operate within predefined bounds, hard ethical constraints must remain invariant, particularly in safety-critical domains. When conflicts cannot be resolved autonomously, human-in-the-loop mechanisms enable stakeholders to arbitrate trade-offs, revise priorities, and resolve ethical tensions at design time or runtime. This process also ensures accountability and traceability by recording decisions and their rationales, thereby supporting auditability, trust, and future reassessment.

\textbf{\underline{Step D.1}: Resolve WFI issues and refine specification.} 
This step addresses decision-making over conflicting requirements to resolve well-formedness issues (WFIs) identified during analysis. It focuses on analyzing trade-offs and defining resolution strategies that preserve system consistency while prioritizing stakeholders' needs and associated values. The fine-grained feedback provided by LEGOS-SLEEC, combined with the goal-oriented requirements model, supports the negotiation of adaptation strategies at the goal level, guiding adaptation decisions while preserving normative properties and functional requirements. 
\textit{\underline{Example}.}  
The situational conflict in Fig.~\ref{fig:situational} illustrates a typical case in which a patient withdraws consent for data processing. Resolving this conflict requires involving both technical and non-technical stakeholders. By analyzing the conflict alongside the goal model, participants can explore strategies that reconcile the consent withdrawal with constraints preventing immediate sensor deactivation. Example strategies include: (i) \emph{Data anonymization}, masking personal identifiers; (ii) \emph{Modified communication protocols}, preventing data transmission to the central hub; and (iii) \emph{Postponed decision-making}, continuing data collection for system stability while suspending its use. These strategies mitigate the conflict while respecting both system constraints and stakeholder values.

\section{Preliminary Evaluation}~\label{sec:evaluation}

To evaluate our approach, we analysed its effectiveness for operationalizing human values at design-time.  We answer this question by evaluating (1) the ability of our approach to capture conflicts between normative, functional and adaptation goals, and (2) how the diagnosis helps the negotiation to solve potential conflicts. 

We conducted a preliminary study using the Body Sensor Network (BSN) system~\cite{BSN:SEAMS2021}, building on the same scenarios explored in~\cite{NormativeRequirements:ICSE2024} and exercised in the reference architecture proposed by~\citet{AUTILI2026112749}. %
Following our approach, high-level requirements were elicited, modeled as goals, refined into tasks, and automatically translated into the SLEEC DSL. To support this process, we implemented and publicly released an extension of LEGOS-SLEEC~\cite{LEGOS_SLEEC_XT} that enables automated translation of goals and tasks into SLEEC rules and purposes. 
We then conducted a negotiation session involving two requirements engineering experts and the BSN system engineer. Participants were provided with both the goal model (Fig.~\ref{fig:BSN}) and the LEGOS-SLEEC diagnosis, enabling them to explore strategies to reconcile the patient's privacy request with a critical technical constraint: physical sensors cannot be deactivated without compromising system integrity. As a result, simply reporting failure or halting monitoring was deemed unacceptable, as emphasized by the BSN engineer, who noted that stopping the sensors would undermine the system's primary functionality. 

The negotiation resulted in agreement on a revised adaptation policy, introducing two strategies triggered when the system enters a partial-tracking state: a \emph{Data Anonymization Strategy}, which protects patient identity by anonymizing collected data, and a \emph{Communication Protocol Adaptation Strategy}, which adjusts how data are transmitted. As a result, the requirement restricting outdoor data tracking after profile adaptation was removed. The goal model was then revised accordingly, and a new set of rules was generated.

Thus, this preliminary evaluation shows that our approach can identify conflicts between normative, functional, and adaptation requirements and support their negotiation during requirements operationalization, thereby answering our research question. 
It also reveals that unresolved obstacles are a major source of situational conflicts. By making obstacle events explicit, our approach provides a concrete starting point for negotiating adaptation policies and exposes sources of uncertainty that must be considered by both technical and non-technical stakeholders. Supplementary material of the case study publicly available~\cite{BSN:LEGOS_SLEEC_XT}.
\section{Conclusion and Future Work}~\label{sec:conclusion}
In this work, we propose \approach: a requirements-driven process for operationalizing human values in ethics-aware self-adaptive systems. Our proof-of-concept implementation and initial case study demonstrate the feasibility of explicitly modeling functional, normative, and adaptation goals, and of analyzing their interactions to detect inconsistencies and value conflicts early. These results suggest that our approach effectively supports requirements for self-adaptive systems, while preserving a formal foundation for satisfiability checking and the detection of value conflicts arising from adaptation. 
Future work will address explicit modeling of multiple sources and dimensions of uncertainty to better anticipate adaptation decisions and improve traceability. We also plan to investigate complementary analysis techniques, such as assurance mechanisms and obstacle-resolution strategies, and to conduct larger empirical studies to assess the efficiency and quality of elicited requirements and adaptation strategies.

\newpage

\balance
\bibliographystyle{ACM-Reference-Format}
\bibliography{references}

\end{document}